\documentclass[footinbib,aps,prb,twocolumn,superscriptaddress,groupedaddress,a4paper]{revtex4}  
\usepackage{graphicx}
\usepackage{dcolumn}
\usepackage{bm}
\usepackage{amssymb}
\usepackage{amsmath}
\usepackage{subfigure}
\usepackage{color}
\usepackage{lscape}
\begin{document}

\widetext
\title{Electronic properties of type-II GaAs$_{1-x}$Sb$_{x}$/GaAs quantum rings\\for applications in intermediate band solar cells}


\author{Reza Arkani}
\affiliation{Tyndall National Institute, University College Cork, Lee Maltings, Dyke Parade, Cork T12 R5CP, Ireland}
\affiliation{Department of Physics, University College Cork, Cork T12 YN60, Ireland}

\author{Christopher A.~Broderick}
\email{c.broderick@umail.ucc.ie} 
\affiliation{Tyndall National Institute, University College Cork, Lee Maltings, Dyke Parade, Cork T12 R5CP, Ireland}
\affiliation{Department of Physics, University College Cork, Cork T12 YN60, Ireland}

\author{Eoin P.~O'Reilly}
\affiliation{Tyndall National Institute, University College Cork, Lee Maltings, Dyke Parade, Cork T12 R5CP, Ireland}
\affiliation{Department of Physics, University College Cork, Cork T12 YN60, Ireland}

\date{\today}


\begin{abstract}

We present a theoretical analysis of the electronic properties of type-II GaAs$_{1-x}$Sb$_{x}$/GaAs quantum rings (QRs), from the perspective of applications in intermediate band solar cells (IBSCs). We outline the analytical solution of Schr\"{o}dinger's equation for a cylindrical QR of infinite potential depth, and describe the evolution of the QR ground state with QR morphology. Having used this analytical model to elucidate general aspects of the electronic properties of QRs, we undertake multi-band \textbf{k}$\cdot$\textbf{p} calculations -- including strain and piezoelectric effects -- for realistic GaAs$_{1-x}$Sb$_{x}$/GaAs QRs. Our \textbf{k}$\cdot$\textbf{p} calculations confirm that the large type-II band offsets in GaAs$_{1-x}$Sb$_{x}$/GaAs QRs provide strong confinement of holes, and further indicate the presence of resonant (quasi-bound) electron states which localise in the centre of the QR. From the perspective of IBSC design the calculated electronic properties demonstrate several benefits, including (i) large hole ionisation energies, mitigating thermionic emission from the intermediate band, and (ii) electron-hole spatial overlaps exceeding those in conventional GaAs$_{1-x}$Sb$_{x}$/GaAs QDs, with the potential to engineer these overlaps via the QR morphology so as to manage the trade-off between optical absorption and radiative recombination. Overall, our analysis highlights the flexibility offered by the QR geometry from the perspective of band structure engineering, and identifies specific combinations of QR alloy composition and morphology which offer optimised sub-band gap energies for QR-based IBSCs.

\end{abstract}


\maketitle


\section{Introduction}
\label{sec:introduction}


Intermediate band solar cells (IBSCs) \cite{Luque_PRL_1997,Okasa_APR_2015} constitute an attractive approach to develop next-generation solar cells, since they offer the potential to significantly exceed the single-junction Shockley-Queisser (detailed balance) photovoltaic efficiency limit. \cite{Shockley_JAP_1961} This is achieved via introduction of an intermediate band (IB) lying energetically within the band gap of a host matrix semiconductor. In the case that the IB is electrically isolated from both the valence band (VB) and conduction band (CB) of the host matrix semiconductor -- via gaps in the density of states -- carrier generation due to absorption of photons having energy less than the band gap $E_{g} = E_{\scalebox{0.7}{\textrm{CB}}} - E_{\scalebox{0.7}{\textrm{VB}}}$ of the host matrix semiconductor can occur via two-step photon absorption (TSPA). In a hole-based IBSC, TSPA proceeds via (i) an electron in the IB being promoted to the CB via absorption of a sub-band gap photon having energy $E_{\scalebox{0.7}{\textrm{H}}} = E_{\scalebox{0.7}{\textrm{CB}}} - E_{\scalebox{0.7}{\textrm{IB}}}$, and (ii) subsequent promotion of the resulting IB hole to the VB via the absorption of a second sub-band gap photon having energy $E_{\scalebox{0.7}{\textrm{L}}} = E_{\scalebox{0.7}{\textrm{IB}}} - E_{\scalebox{0.7}{\textrm{VB}}}$. The presence of the IB therefore enhances the photocurrent generated by a single-junction solar cell at fixed illumination, by allowing absorption of photons having energy $< E_{g}$, while the electrical isolation of the IB from the host matrix CB and VB ensures that the open-circuit voltage $V_{\scalebox{0.7}{\textrm{OC}}}$ is determined by $E_{g}$ rather than being limited by the sub-band gap energies $E_{\scalebox{0.7}{\textrm{L}}}$ and $E_{\scalebox{0.7}{\textrm{H}}}$. \cite{Luque_PRL_1997,Luque_AM_2010}

Due to their promise of high efficiency -- with a predicted detailed balance limit of 63.8\% under concentrated illumination \cite{Luque_AM_2010,Luque_PRL_1997,Okasa_APR_2015} -- IBSCs have attracted significant research interest. However, despite the simplicity of the underlying concepts, practical realisation of IBSCs has proved extremely challenging. \cite{Marti_JPE_2013,Okasa_APR_2015} Practical efforts to realise IBSCs have to date centred primarily on two approaches to introduce an IB into the band gap of a host matrix semiconductor, relying on (i) a bound electron or hole ground state in a three-dimensional quantum confined heterostructure possessing a discrete density of states, e.g.~in a quantum dot (QD), or (ii) an IB formed via incorporation of a small concentration of a substitutional impurity, e.g.~in a highly-mismatched alloy (HMA) such as dilute nitride GaN$_{x}$As$_{y}$P$_{1-x-y}$. \cite{Kudraweic_PRA_2014} The performance of QD-IBSCs has been limited by a combination of poor sub-band gap absorption, which limits short-circuit current density $J_{\scalebox{0.7}{\textrm{SC}}}$, and short radiative lifetimes $\tau_{\scalebox{0.7}{\textrm{rad}}}$ for carriers occupying IB states, resulting in loss of carriers from the IB via radiative recombination and so reducing $V_{\scalebox{0.7}{\textrm{OC}}}$. Also, the presence of insufficiently large band offsets in QDs can lead to loss of carriers via thermionic emission (e.g.~from the IB to CB in an electron-based QD-IBSC). \cite{Ramiro_IEEEJP_2015} The use of localised impurity states in HMA-IBSCs also presents fundamental issues, increasing losses associated with non-radiative (Shockley-Read-Hall) recombination of carriers occupying IB states at defect sites, thereby limiting carrier extraction and degrading overall efficiency.


As a result, research efforts to realise IBSCs have increasingly shifted away from conventional platforms such as type-I QDs, and towards novel materials and heterostructures whose electronic and optical properties offer the potential to overcome the aforementioned limitations. In particular, there is increasing interest in heterostructures having type-II band alignment, \cite{Kechiantz_N_2007,Tayagaki_APL_2012,Kechiantz_PP_2015,Takeshi_APL_2016} due to their potential to suppress radiative losses as a consequence of the intrinsically large radiative lifetimes associated with their bound eigenstates (resulting from reduced electron-hole spatial overlaps), \cite{Cuadra_PE_2002,Nishikawa_APL_2012} as well as reduced intra-band carrier relaxation via non-radiative (Auger) recombination. \cite{Tomic_APL_2011,Tomic_APL_2013} Much of the work to date on type-II QD-IBSC systems has centred on InAs/GaAs$_{1-x}$Sb$_{x}$ QDs -- an electron-based IBSC, in which the IB is formed by the lowest energy bound electron states in the InAs dot, and holes are localised in a GaAs$_{1-x}$Sb$_{x}$ quantum well buffer layer -- where increases in QD uniformity and number density have been reported in epitaxial growth of vertical QD stacks. \cite{Ban_APL_2010,Liu_SEMSC_2012,Hatch_OE_2014,Cheng_SEMSC_2016} Experimental investigations have revealed increased $J_{\scalebox{0.7}{\textrm{SC}}}$ compared to conventional type-I InAs/GaAs QDs, \cite{Liu_SEMSC_2012} in line with the theoretically predicted increase in the radiative lifetime of IB states in these structures. \cite{Tomic_APL_2013} While these promising results highlight the potential of type-II heterostructures as candidate IBSCs, the InAs/GaAs(Sb) system suffers from a relatively low CB offset, leading to carrier leakage via thermionic emission from the IB at room temperature. \cite{Cheng_SEMSC_2016} Furthermore, the sub-band gaps in InAs/GaAs(Sb) QDs are far from the optimum values $E_{\scalebox{0.7}{\textrm{H}}} = 0.97$ eV and $E_{\scalebox{0.7}{\textrm{L}}} = 0.45$ eV required to maximise the calculated detailed balance efficiency for an IBSC based on a GaAs host matrix semiconductor. \cite{Wang_IETO_2014}

What is therefore required is to broaden investigations to additional type-II heterostructure systems, focusing on the ability to engineer the electronic properties so as to reliably tune the energy and character of the IB states in order to maximise overall efficiency. One such class of novel type-II heterostructures are GaAs$_{1-x}$Sb$_{x}$/GaAs quantum rings (QRs). These hole-based IBSCs -- in which the IB is formed by the highest energy bound hole eigenstate in the QR -- have attracted increasing attention due not only to their type-II band alignment, but also to their large band offsets, which are expected to mitigate carrier loss via thermionic emission from the IB. Experimental analysis of prototype IBSCs based on type-II GaAs$_{1-x}$Sb$_{x}$/GaAs vertical QR stacks has revealed several promising properties compared to conventional QD-IBSCs, including (i) enhanced TSPA and external quantum efficiency, \cite{Wagener_JAP_2014_2,Shoji_AIPA_2017} (ii) reduced losses via radiative recombination, leading to improved carrier extraction and overall efficiency, \cite{Hwang_JAP_2012} and (iii) recovery of $V_{\scalebox{0.7}{\textrm{OC}}}$ under concentrated illumination. \cite{Tsai_OE_2014,Fujita_PP_2015,Montesdeoca_SEMSC_2018}


Despite numerous and ongoing experimental investigations of type-II GaAs$_{1-x}$Sb$_{x}$/GaAs QRs for IBSC applications, there is little detailed information available in the literature regarding their electronic properties from a theoretical perspective. Here, we present a combined analytical and numerical analysis of the electronic properties of GaAs$_{1-x}$Sb$_{x}$/GaAs QRs and demonstrate that minor changes in morphology, compatible with established epitaxial growth, can be exploited to tune the QR hole ground state (IB) to an optimum energy to maximise IBSC efficiency. Using an analytical analysis -- based on a solution of the time-independent Schr\"{o}dinger equation for a cylindrical QR of infinite potential depth -- we highlight that the QR geometry offers significant flexibility, compared to the conventional QD geometry, to engineer the valence band (VB) structure of GaAs$_{1-x}$Sb$_{x}$/GaAs structures for IBSC applications. Our numerical calculations -- based on a multi-band \textbf{k}$\cdot$\textbf{p} Hamiltonian, and including full strain and piezoelectric effects -- corroborate this finding. We highlight that type-II GaAs$_{1-x}$Sb$_{x}$/GaAs QRs are ideally suited to IBSC applications, due not only to their intrinsically large radiative lifetimes, but also due to their large VB offsets, which can be expected to mitigate thermionic emission of holes from the IB. Furthermore, the strong confinement of the highest energy hole (IB) states in these QRs is expected to mitigate the impact of miniband formation via electronic coupling between QRs in high-density stacks, which has been demonstrated to lead to a closing of the gap in the density of states between the IB and CB in electron-based InAs/GaAs QD stacks. \cite{Tomic_PRB_2010}

We undertake a numerical optimisation of the QR morphology, by varying the QR dimensions and alloy composition, to identify structures which allow to achieve optimum IB energy so as to maximise IBSC efficiency. Our results confirm the potential of GaAs$_{1-x}$Sb$_{x}$/GaAs QRs for IBSC applications, and provide guidelines for the growth of suitable structures for prototype IBSCs.


The remainder of this paper is organised as follows. In Sec.~\ref{sec:theoretical_model} we outline the theoretical models we have developed to calculate the electronic properties of GaAs$_{1-x}$Sb$_{x}$/GaAs QRs, describing analytical and numerical approaches in Secs.~\ref{sec:theoretical_model_analytical} and~\ref{sec:theoretical_model_numerical} respectively. OWe present our results in Sec.~\ref{sec:results}, beginning in Sec.~\ref{sec:results_analytical} with an analysis of the QR ground state obtained from the analytical model. In Sec.~\ref{sec:results_numerical} we describe the electronic structure of real GaAs$_{1-x}$Sb$_{x}$/GaAs QRs, based on numerical multi-band \textbf{k}$\cdot$\textbf{p} calculations. Finally, in Sec.~\ref{sec:conclusions} we summarise and conclude.


\section{Theoretical model}
\label{sec:theoretical_model}

In this section we describe the theoretical approaches we have applied to investigate the electronic properties of GaAs$_{1-x}$Sb$_{x}$/GaAs QRs. We begin in Sec.~\ref{sec:theoretical_model_analytical} with a description of the analytical solution of Schr\"{o}dinger's equation for a QR of infinite potential depth, and in Sec.~\ref{sec:theoretical_model_numerical} describe numerical strain relaxation and multi-band \textbf{k}$\cdot$\textbf{p} calculations for realistic QRs.


\subsection{Analytical: solution of Schr\"{o}dinger's equation}
\label{sec:theoretical_model_analytical}

\begin{figure*}[ht!]
	\includegraphics[width=1.00\textwidth]{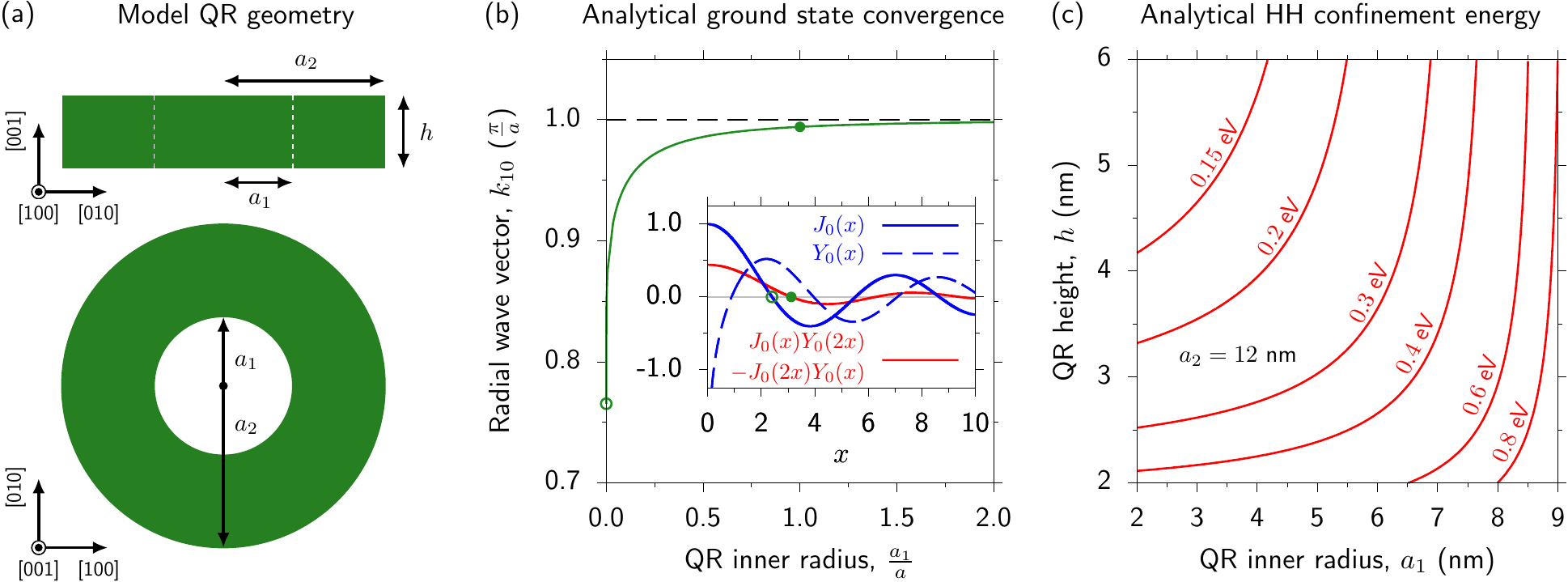}
	\caption{(a) Schematic illustration of the QR geometry considered in this work, viewed side-on from the [100] direction (top) and top-down from the [001] direction (bottom). The cylindrical QR has inner and outer radii $a_{1}$ and $a_{2}$, radial thickness $a = a_{2} - a_{1}$, and height $h$ along the growth direction. (b) Variation of the ground state radial wave vector $k_{10}$, computed as the first root of Eq.~\eqref{eq:quantum_ring_transcendental} for $m = 0$, as a function of the inner radius $a_{1}$ for a QR of fixed radial thickness $a$. As $a_{1}$ increases as a proportion of $a$, $k_{10}$ rapidly converges to a value of $\frac{ \pi }{ a }$ (dashed black line). The inset shows the Bessel functions $J_{0} (x)$ (solid blue line) and $Y_{0} (x)$ (dashed blue line), as well as the left-hand side of Eq.~\eqref{eq:quantum_ring_transcendental} for $\rho = 2$ (i.e.~$a_{1} = a$). (c) Contour map showing the confinement energy associated with the HH ground state in a GaSb/GaAs QR having outer radius $a_{2} = 12$ nm, calculated via Eqs.~\eqref{eq:quantum_ring_transcendental} and Eq.~\eqref{eq:quantum_ring_ground_state_energy}, as a function of the QR inner radius $a_{1}$ and height $h$.}
  \label{fig:analytical}
\end{figure*}

Our analytical analysis starts via the solution of the time-independent Schr\"{o}dinger equation for the eigenstates of an idealised, [001]-oriented cylindrical QR of infinite potential depth. The QR is taken to have inner and outer radii $a_{1}$ and $a_{2}$ in the plane perpendicular to the [001] direction, and height $h$ along [001]. A schematic illustration of the QR geometry is shown in Fig.~\ref{fig:analytical}(a). Choosing the origin of a cylindrical polar coordinate system $( r, \phi, z )$ to lie at the centre of the base of the QR, the potential -- which is axially symmetric about the [001], or $z$, direction and hence independent of $\phi$ -- is

\begin{equation}
	V (r, h) = \left\{
		\begin{array}{lr}
			0 \; , & a_{1} < r < a_{2}~\textrm{and}~0 < z < h \\
			\infty, & \textrm{otherwise}
		\end{array}
		\right. \, .
	\label{eq:quantum_ring_potential}
\end{equation}


We proceed via separation of variables by writing the QR eigenstates $\psi_{lmn} ( r, \phi, z ) = R_{lm} ( r ) \Phi_{m} ( \phi ) Z_{n} ( z )$, yielding three separable differential equations (one in each of the radial, polar and longitudinal coordinates $r$, $\phi$ and $z$). The solution of the $\phi$ and $z$ equations are as in the conventional cylindrical QD of infinite potential depth (corresponding here to $a_{1} = 0$). The solution in the polar ($\phi$) direction is trivial due to the axial symmetry of the potential about the $z$ direction, with $\Phi_{m} ( \phi ) = \frac{ 1 }{ \sqrt{ 2 \pi } } \, e^{ i m \phi }$ for integer $m$. Along the $z$ direction the eigenstates are those of the infinite square well

\begin{equation}
    Z_{n} ( z ) = \sqrt{ \frac{ 2 }{ h } } \sin \left( \frac{ n \pi z }{ h } \right) \, .
    \label{eq:general_solution_longitudinal}
\end{equation}

Within the QR $V ( r, z ) = 0$ and the radial equation reduces to the (cylindrical) Bessel equation, the general solution of which is $R_{lm} ( r ) = A_{lm} J_{m} ( kr ) + B_{lm} Y_{m} ( kr )$, where $J_{m} ( kr )$ and $Y_{m} ( kr )$ are respectively the Bessel functions of the first and second kind. Here, $l$ is a positive integer which indexes the (discrete) allowed values $k_{lm}$ of the radial wave vector $k$. To this point, the general solution is identical to that of a cylindrical QD of infinite potential depth. The difference in the QR case arises due to the presence of the central barrier for $r \leq a_{1}$. Due to the presence of the central barrier we seek radial wave functions satisfying the boundary conditions $R_{lm} ( a_{1} ) = 0$ and $R_{lm} ( a_{2} ) = 0$. Applying the first of these conditions allows us to solve for $B_{lm}$ in terms of $A_{lm}$, giving

\begin{equation}
    R_{lm} ( r ) = A_{lm} \left( J_{m} ( kr ) - \frac{ J_{m} ( k a_{1} ) }{ Y_{m} ( k a_{1} ) } \, Y_{m} ( kr ) \right) \, ,
    \label{eq:radial_wave_function}
\end{equation}

\noindent
where the constant $A_{lm}$ can be determined via normalisation. Applying the second boundary condition then yields the transcendental equation

\begin{equation}
    J_{m} ( k a_{1} ) \, Y_{m} ( \rho k a_{1} ) - J_{m} ( \rho k a_{1} ) \, Y_{m} ( k a_{1} ) = 0 \, ,
    \label{eq:quantum_ring_transcendental}
\end{equation}

\noindent
where we have defined $\rho = \frac{ a_{2} }{ a_{1} }$ as the ratio of the outer and inner QR radii (so that $\rho k a_{1} = k a_{2}$). The wave vector $k_{lm}$ associated with the radial wave function $R_{lm} ( r )$ is then, for a given value of $m$, determined via the $l^{\scalebox{0.7}{\textrm{th}}}$ root $k_{lm} a_{1}$ of Eq.~\eqref{eq:quantum_ring_transcendental}. Defining $a = a_{2} - a_{1}$ as the radial thickness of the QR, we note that Eq.~\eqref{eq:radial_wave_function} reduces to $R_{lm} ( r ) = A_{lm} J_{m} ( kr )$ in the limit $a_{1} \to 0$ ($a_{2} \to a$), with Eq.~\eqref{eq:quantum_ring_transcendental} correspondingly reducing to $J_{m} ( ka ) = 0$, yielding the well-known solution of Schr\"{o}dinger's equation for a cylindrical QD of radius $a$ as the QR is transformed into a QD via the removal of the central potential barrier.

The energies of the QR eigenstates are given by the sum of the in- and out-of-plane contributions

\begin{equation}
    E_{lmn} = \frac{ \hbar^{2} }{ 2 m_{0} } \left( \frac{ k_{lm}^{2} }{ m_{\perp}^{\ast} } + \frac{ \pi^{2} n^{2} }{ m_{\parallel}^{\ast} h^{2} } \right) \, ,
    \label{eq:quantum_ring_ground_state_energy}
\end{equation}

\noindent
where $m_{0}$ is the free electron mass, and where $m_{\parallel}^{\ast}$ and $m_{\perp}^{\ast}$ are respectively the (relative) effective mass parallel and perpendicular to the [001] growth direction. The QR ground state is obtained for quantum numbers $( l, m, n ) = ( 1, 0, 1 )$.


\subsection{Numerical: multi-band \textbf{k}$\cdot$\textbf{p} calculations}
\label{sec:theoretical_model_numerical}

Our numerical analysis of the electronic properties of GaAs$_{1-x}$Sb$_{x}$/GaAs QRs is based on multi-band \textbf{k}$\cdot$\textbf{p} calculations. \cite{Arkani_NUSOD_2019} We employ a supercell approach, by embedding a cylindrical GaAs$_{1-x}$Sb$_{x}$/GaAs QR in a GaAs matrix, and relax the supercell by minimising the total elastic energy with respect to the components $\epsilon_{ij}$ of the strain tensor to obtain the strain fields $\epsilon_{ij} ( \textbf{r} )$. The QR VB eigenstates are computed using a strain-dependent 6-band \textbf{k}$\cdot$\textbf{p} Hamiltonian -- i.e.~the Luttinger-Kohn VB Hamiltonian, including Bik-Pikus strain-related terms -- which explicitly treats heavy-hole (HH), light-hole (LH) and spin-split-off (SO) VB states. The QR CB eigenstates are computed using a strain-dependent 1-band (effective mass) Hamiltonian. We note that the separate treatment of the CB and VB eigenstates has been chosen to circumvent the emergence of spurious solutions in full 8-band \textbf{k}$\cdot$\textbf{p} calculations, which arise in the presence of large plane wave cut-off energies due to the strong inter-band coupling present in narrow-gap GaSb. Both the 6-band VB and 1-band CB calculations explicitly include the strain-induced piezoelectric potential, computed to second order for a given structure using the relaxed strain fields. The QR band offsets are computed firstly for an unstrained structure using the model solid theory -- assuming a natural (unstrained) VB offset of 0.58 eV between GaSb and GaAs \cite{Wei_APL_1998,Hinuma_PRB_2014} -- with the position-dependent band edge energy (confining potential) profiles then computed via direct diagonalisation of the strain-dependent bulk 6- and 1-band \textbf{k}$\cdot$\textbf{p} Hamiltonians at each real space grid point in the supercell.

Our numerical calculations have been implemented using a plane wave (reciprocal space) approach, via the S/Phi/nX software library. \cite{Marquardt_CPC_2010} We employ [001]-oriented supercells of size of 50 nm $\times$ 50 nm $\times$ 14 nm, and a plane wave cut-off energy equivalent to a real space resolution of 0.2 nm in each of the $x$, $y$ and $z$ directions. This choice of plane wave cut-off was validated by examining the convergence of the bound carrier eigenstate energies, which were found to vary by $<1$ meV with respect to further increases in the size of the plane wave basis set. All material parameters used in our calculations -- including lattice and elastic constants, band gaps, VB spin-orbit splitting energies and Luttinger parameters, electron effective masses, and CB and VB edge deformation potentials -- are taken from Ref.~\onlinecite{Vurgaftman_JAP_2001}, with the exception of the first and second order piezoelectric coefficients, which are taken from Ref.~\onlinecite{Bester_PRB_2006}. All calculations are performed at temperature $T = 300$ K.


\section{Results}
\label{sec:results}

In this section we present the results of our theoretical analysis, beginning in Sec.~\ref{sec:results_analytical} with a description of trends in the electronic properties of GaAs$_{1-x}$Sb$_{x}$/GaAs QRs based on the analytical treatment described in Sec.~\ref{sec:theoretical_model_analytical}. Specifically, we describe (i) the rapid convergence of the ground state radial wave vector $k_{10}$ in a QR having fixed radial thickness $a = a_{2} - a_{1}$, allowing the ground state energy in a realistic QR to be easily and accurately estimated, and (ii) trends in the HH confinement energy in realistic GaAs$_{1-x}$Sb$_{x}$/GaAs QRs. In Sec.~\ref{sec:results_numerical} we present the results of our numerical calculations, including the strain fields and band offsets (confining potentials) in relaxed QRs, as well as the electronic properties obtained from multi-band \textbf{k}$\cdot$\textbf{p} calculations, allowing morphologies supporting optimised IBSC sub-band gaps to be identified.


\subsection{Analytical: ground state radial wave vector convergence and QR HH confinement energy}
\label{sec:results_analytical}


We begin by demonstrating a useful convergence property of the ground state radial wave vector $k_{10}$, obtained from the first ($l = 0$) root of Eq.~\eqref{eq:quantum_ring_transcendental} for $m = 0$. Specifically, for a QR having radial thickness $a = a_{2} - a_{1}$, $k_{10}$ converges rapidly to $\frac{ \pi }{ a }$ with increasing inner radius $a_{1}$. This provides a useful approximation, $k_{10} \approx \frac{ \pi }{ a }$, which can be used to estimate the radial contribution to the QR ground state energy (cf.~Eq.~\eqref{eq:quantum_ring_ground_state_energy}).

To see that this is the case we note that, for large $x$, $J_{0} (x) \approx \sqrt{ \frac{ 2 }{ \pi x } } \cos \left( x - \frac{ \pi }{ 4 } \right)$ and $Y_{0} (x) \approx \sqrt{ \frac{ 2 }{ \pi x } } \sin \left( x - \frac{ \pi }{ 4 } \right)$, so that the $m = 0$ radial wave function can be written in the form

\begin{equation}
    R_{l0} (r) \approx C_{l0} \sqrt{ \frac{ 2 }{ \pi k r } } \cos \left( kr + \theta \right) \, ,
    \label{eq:radial_wave_function_limit}
\end{equation}

\noindent
for some phase $\theta$ and normalisation constant $C_{l0}$. This approximate expression for $R_{l0} (r)$ satisfies the boundary conditions $R_{l0} ( a_{1} ) = 0$ and $R_{l0} ( a_{2} ) = 0$ for radial wave vectors $k_{l0} \sim \frac{ l \pi }{ a }$. We therefore deduce that, for a QR of fixed radial thickness $a$, as the inner radius $a_{1}$ increases, the first root $k_{10}$ of Eq.~\eqref{eq:quantum_ring_transcendental} should converge to $\frac{ \pi }{ a }$.

To verify that this is the case we have computed the first root $k_{10} a_{1}$ of Eq.~\eqref{eq:quantum_ring_transcendental} as a function of $a_{1}$ for a QR having fixed thickness $a$ (= 11.5 nm). The results of this analysis are summarised in Fig.~\ref{fig:analytical}(b), where the solid green line shows the calculated variation of $k_{10}$ (in units of $\frac{ \pi }{ a }$) as a function of $\frac{ a_{1} }{ a }$ ($= \frac{ 1 }{ \rho - 1 }$). The inset to Fig.~\ref{fig:analytical}(b) shows the functions $J_{0} (x)$ (solid blue line) and $Y_{0} (x)$ (dashed blue line), as well as the left-hand side of Eq.~\eqref{eq:quantum_ring_transcendental} for $m = 0$ and $\rho = 2$ ($a_{2} = 2 a_{1}$, solid red line). Examining Fig.~\ref{fig:analytical}(b), we in fact note \textit{rapid} convergence of $k_{10}$ to $\frac{ \pi }{ a }$ with increasing $\frac{ a_{1} }{ a }$. For $a_{1} = 0$ -- i.e.~for a cylindrical QD of radius $a$ -- $k_{10}$ is given by the first root of $J_{0} ( ka ) = 0$. This solution, highlighted by the open green circles in Fig.~\ref{fig:analytical}(b) and its inset, is $k_{10} a \approx 2.4048$ (the first zero of $J_{0} (ka)$), so that $k_{10} \approx 0.7655 \times \frac{ \pi }{ a }$. As $\frac{ a_{1} }{ a }$ increases due to the inclusion of the central potential barrier to form a QR, $k_{10}$ then rapidly approaches $\frac{ \pi }{ a }$. For example, for $\frac{ a_{1} }{ a } = 1$ ($\rho = 2$), which corresponds well to real QR dimensions, we compute $k_{10} \approx 0.9941 \times \frac{ \pi }{ a }$.

Generally, we find $k_{10} = \frac{ \pi }{ a }$ to be an excellent approximation to the ground state radial wave vector for $\frac{ a_{1} }{ a } \gtrsim 1$ -- i.e.~for $a_{1} \gtrsim \frac{ a_{2} }{ 2 }$ or, equivalently, $\rho \lesssim 2$. For $a = 11.5$ nm we note that $\rho = 2$ corresponds to an outer QR radius $a_{2} = 2 a_{1} = 23$ nm, dimensions typical of epitaxially grown GaAs$_{1-x}$Sb$_{x}$/GaAs QRs, which will be analysed in further detail in Sec.~\ref{sec:results_numerical}. This rapid convergence of $k_{10}$ then provides a simple and reliable approach to estimate the radial contribution to the QR ground state energy, circumventing the requirement to numerically compute the roots of Eq.~\eqref{eq:quantum_ring_transcendental}, with the overall accuracy of the corresponding estimate of the ground state energy $E_{101}$ being better for small QR aspect ratios $\frac{ h }{ 2 a_{2} }$ (where the radial component contributes a small proportion of the total confinement energy, cf.~Eq.~\eqref{eq:quantum_ring_ground_state_energy}).


We have used Eq.~\eqref{eq:quantum_ring_ground_state_energy} to estimate the confinement energy associated with the HH ground state in a GaSb/GaAs QR. To do so we set $m_{\parallel}^{\ast} = ( \gamma_{1} - 2 \gamma_{2} )^{-1}$ and $m_{\perp}^{\ast} = ( \gamma_{1} + 2 \gamma_{2} )^{-1}$, which are the bulk HH VB edge effective masses admitted by the 6-band Luttinger-Kohn Hamiltonian, \cite{Luttinger_PR_1955,Vurgaftman_JAP_2001} where $\gamma_{1}$ and $\gamma_{2}$ are the VB Luttinger parameters. Following Ref.~\onlinecite{Vurgaftman_JAP_2001} we set $\gamma_{1} = 13.4$ and $\gamma_{2} = 4.7$ and respectively obtain $m_{\parallel}^{\ast} = 0.250$ and $m_{\perp}^{\ast} = 0.044$ for the (relative) HH effective masses parallel and perpendicular to the [001] direction in GaSb. The results of our calculations are summarised in Fig.~\ref{fig:analytical}(c), in which the solid red lines are contours of constant HH ground state (confinement) energy as a function of inner radius $a_{1}$ and height $h$, for a QR having fixed outer radius $a_{2} = 12$ nm. We note from these results the flexibility offered by the QR geometry from the perspective of band structure engineering for IBSC applications: the HH confinement energy can readily be tuned across a broad range via relatively minor adjustments in QR morphology, allowing the IB energy to be tuned in a hole-based QR-IBSC. This provides distinct advantages compared to, e.g., equivalent GaAs$_{1-x}$Sb$_{x}$/GaAs QDs, since the QR inner radius $a_{1}$ provides an additional parameter by which the electronic properties can be tuned. The realities of epitaxial growth do not in general allow $a_{1}$ to be fine tuned independently of the other QR dimensions $a_{2}$ and $h$, the relationships between which are in practice determined in large part by the (Stranski-Krastanov) strain relaxation mechanism that drives QR formation. However, it is generally observed that Stranski-Krastanov QR formation tends to fix $a_{2}$, with the inner radius $a_{1}$ then depending largely on growth rate. Correspondingly, QR heights are generally found to be in the range $h = 3 \pm 2$ nm. \cite{Khan_EMC_2016} Despite incomplete control over the precise morphology of individual QRs during epitaxial growth, we emphasise that GaAs$_{1-x}$Sb$_{x}$/GaAs QRs offer additional benefits for IBSC applications due to the nature of carrier localisation within these structures. To elucidate these properties requires a more detailed, quantitative analysis of the electronic properties, to which we now turn our attention.


\subsection{Numerical: strain, band offsets, carrier localisation and IBSC sub-band gaps}
\label{sec:results_numerical}


While the analytical treatment of Secs.~\ref{sec:theoretical_model_analytical} and~\ref{sec:results_analytical} provides useful insight into some of the general features of the electronic properties of QRs, this approach neglects several key factors which play an important role in determining the nature of the electronic properties in a real semiconductor QR. Firstly, the large lattice mismatch which drives QR formation -- 7.2\% in the case of GaSb/GaAs -- produces large, strongly position dependent strain fields, which impact the electronic properties directly (as well as via the associated strain-induced piezoelectric potential). Secondly, the natural type-II band offsets between GaSb and GaAs produce confining potentials which are markedly different in nature for electrons and holes. Thirdly, the analytical treatment described above neglects band hybridisation effects, which can be expected to play a role in determining precise the nature of QR eigenstates in the presence of strain, quantum-confinement, spin-orbit coupling and narrow band gap, all of which are present in real GaAs$_{1-x}$Sb$_{x}$/GaAs QRs. In order to quantitatively understand the QR electronic structure we have therefore undertaken multi-band \textbf{k}$\cdot$\textbf{p} calculations, based on the formalism described in Sec.~\ref{sec:theoretical_model_numerical}. We provide here an overview of the results of these calculations, elucidating the electronic properties of GaAs$_{1-x}$Sb$_{x}$/GaAs QRs and identifying optimised QR morphologies providing electronic properties well-suited to IBSC applications.

\begin{figure*}[ht!]
	\includegraphics[width=1.00\textwidth]{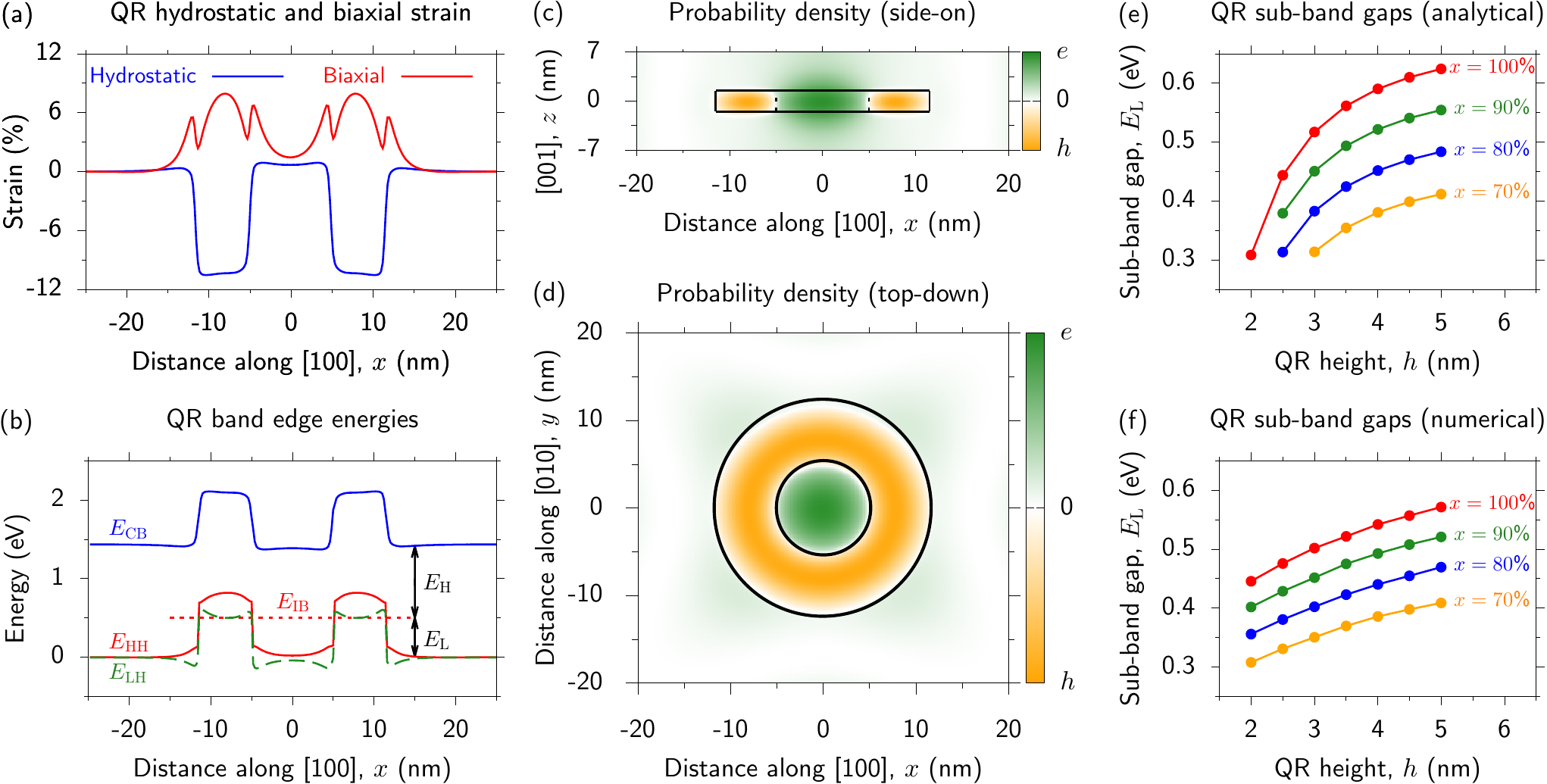}
	\caption{(a) Linescan along the [100] in-plane direction, through the centre of the QR, showing the calculated hydrostatic (solid blue line) and biaxial (solid red line) components of the strain field for a GaSb/GaAs QR having $a_{1} = 5$ nm, $a_{2} = 11.5$ nm and $h = 3$ nm. (b) Linescan along the [100] in-plane direction, through the centre of the QR, showing the calculated bulk band edge energies (band offsets) for the CB (solid blue line), and light-hole (LH; dashed green line) and heavy-hole (HH; solid red line) VBs, for the same QR as in (a). The dashed red line shows the energy of the highest energy bound hole state $h1$ -- i.e.~the intermediate band -- at energy $E_{\protect\scalebox{0.7}{\textrm{IB}}}$, separated from the CB and VB edges by the sub-band gaps $E_{\protect\scalebox{0.7}{\textrm{H}}}$ and $E_{\protect\scalebox{0.7}{\textrm{L}}}$ respectively. (c) Side-on view, along the [010] direction, of the probability density associated with the lowest energy quasi-bound electron (green) and highest energy bound hole (orange) states in the same QR as in (a). (d) Top-down view, along the [001] direction, of the probability density associated with the same two states as in (c). (e) Lower energy sub-band gap $E_{\protect\scalebox{0.7}{\textrm{L}}}$ as a function of QR height $h$ for GaAs$_{1-x}$Sb$_{x}$/GaAs QRs having inner and outer radii $a_{1} = 5$ nm and $a_{2} = 11.5$ nm, estimated based on Eqs.~\eqref{eq:quantum_ring_transcendental} and~\eqref{eq:quantum_ring_ground_state_energy}. (f) As in (e), but with the values of $E_{\protect\scalebox{0.7}{\textrm{L}}}$ obtained from full numerical multi-band \textbf{k}$\cdot$\textbf{p} calculations.}
  \label{fig:numerical}
\end{figure*}

For our \textbf{k}$\cdot$\textbf{p} calculations we begin by considering an exemplar GaSb/GaAs QR having inner radius $a_{1} = 5$ nm, outer radius $a_{2} = 11.5$ nm and height $h = 3$ nm (dimensions typical of epitaxially grown QRs \cite{Khan_EMC_2016}). The solid blue and red lines in Fig.~\ref{fig:numerical}(a) respectively show linescans -- taken through the centre of the QR along the [100] direction -- of the calculated hydrostatic and biaxial components $\epsilon_{\scalebox{0.7}{\textrm{hyd}}} = \textrm{Tr}( \epsilon )$ and $\epsilon_{\scalebox{0.7}{\textrm{bia}}} = \epsilon_{zz} - \frac{1}{2} ( \epsilon_{xx} + \epsilon_{yy} )$ of the strain fields in the structure. The hydrostatic strain reaches values as low as $-10.5$\% within the GaSb regions, reflecting that the QR is under significant compressive strain when grown epitaxially on GaAs. We note that strain relaxation in the structure acts to place the central GaAs region of the QR under a minor amount ($\lesssim 1$\%) of tensile strain, due to its being surrounded in the plane perpendicular to [001] by GaSb material having a larger lattice constant. The calculated biaxial strain resembles that associated with a cylindrical QD, \cite{Andreev_JAP_1999} varying strongly at the interfaces with the surrounding barrier material, albeit with additional features due to the presence of the central GaAs barrier. We note that the strain-induced piezoelectric potential associated with this structure attains a maximum value of $\approx 17$ meV throughout the calculational supercell, and has minimal impact on the electronic properties.


Figure~\ref{fig:numerical}(b) shows a linescan of the band edge energies for the same GaSb/GaAs QR, calculated using the strain fields shown (in part) in Fig.~\ref{fig:numerical}(a). The band edge energy (confining potential) profiles are shown -- using solid blue, solid red and dashed green lines for the CB, HH VB and LH VB edges, respectively -- as a linescan along the [100] direction, through the centre of the QR. We firstly note the presence of large type-II band offsets in the QR. There exist strong confining potentials for holes within the GaSb region, reaching a maximum depth of $\approx 820$ meV for HH states, while electrons are excluded from the GaSb region by a CB potential barrier having a maximum height of $\approx 670$ meV. Due to the compressive strain in the GaSb region, the HH VB edge is pushed higher in energy than the LH VB edge, resulting in a larger confining potential for HH states and ensuring that the highest energy hole bound state -- i.e.~the IB in a hole-based GaAs$_{1-x}$Sb$_{x}$/GaAs QR-IBSC -- is primarily HH-like. We therefore expect that hole states will be strongly confined within the GaSb region, experiencing strong annular localisation in the plane perpendicular to the [001] direction, as well as strong localisation along the [001] direction. We also note that the aforementioned minor tensile strain in the central barrier of the QR tends to reduce slightly the CB edge energy in the centre of the QR. This, combined with the presence of a confining potential for electrons in the plane perpendicular to the [001] direction, suggests the possibility of quasi-localised electron states residing in the central barrier region.

These expected trends in carrier localisation are verified by our calculated probability densities, shown in Figs.~\ref{fig:numerical}(c) and~\ref{fig:numerical}(d), which respectively show cross-sections through the centre of the QR of the calculated electron (green) and hole (orange) probability densities in the plane perpendicular to the [010] and [001] directions. We observe very strong localisation of the highest energy hole bound state within the QR, with only minimal penetration of the hole probability density into the GaSb region. We also note the presence of a resonant electron state within the central GaAs barrier of the QR, which our calculations identify as a consequence of the type-II band offsets  (cf.~Fig.~\ref{fig:numerical}(b)). While this electron state is strongly localised in the plane of the QR, it is less strongly localised along [001], due to the absence of a strongly confining CB potential along that direction.

We note that the quasi-localised electron state for which the probability density is shown in Figs.~\ref{fig:numerical}(c) and~\ref{fig:numerical}(d) was not the lowest energy electron state identified in our calculation: it was in fact the sixth lowest energy electron state. The type-II band alignment in this GaSb/GaAs QR results in a large range of delocalised electron states lying at and above the GaAs CB edge in energy, which are characterised by their probability density being excluded from the central region of the QR in the plane of the QR. (The precise number of delocalised states appearing in a given energy range in a numerical calculation depends on a combination of the supercell dimensions and the size of the plane wave basis set employed.) The quasi-bound electron state identified in our analysis lies only 31 meV above the GaAs CB edge in energy and, given its higher spatial overlap with the highest energy bound hole state compared to the delocalised CB edge states described above, transitions involving this state can be expected to contribute significantly to the band edge optical absorption. From the perspective of IBSC design and optimisation, the presence of this unusual quasi-bound electron state is particularly appealing. By varying the QR morphology the in-plane confinement of this electron state can be engineered so as to control the spatial overlap with the highest energy bound hole state. This overlap governs a key trade-off in an IBSC, between (i) the generation rate associated with carriers occupying IB states (i.e.~the generation of holes in the VB via absorption of photons having energy $\geq E_{\scalebox{0.7}{\textrm{L}}}$), and (ii) the radiative lifetime for recombination involving electron and hole states (i.e.~recombination of quasi-bound CB electrons with bound IB holes via emission of photons having energy $> E_{\scalebox{0.7}{\textrm{H}}}$). Furthermore, the weak electron localisation along [001] suggests that photo-generated electrons can be readily collected from the QRs in a real IBSC structure, with hole extraction proceeding via excitation of holes from the IB to the GaAs VB via TSPA. Finally, the large VB offsets in these QRs result in very large ionisation energies of $\approx 0.5$ eV for IB hole states. This ionisation energy is far in excess of the thermal energy in the range of temperatures relevant to IBSC operation, and so electrical leakage of carriers via thermionic emission from the IB should be effectively suppressed in these heterostructures.


For this GaSb/GaAs QR we calculate that the highest energy bound hole state -- the energy of which is denoted by a horizontal dotted red line in Fig.~\ref{fig:numerical}(b) -- lies 502 meV above the GaAs barrier VB edge, corresponding to IBSC sub-band gaps $E_{\scalebox{0.6}{\textrm{L}}} = 0.502$ eV and $E_{\scalebox{0.6}{\textrm{H}}} = E_{g} (\textrm{GaAs}) - E_{\scalebox{0.6}{\textrm{L}}} = 0.922$ eV at temperature $T = 300$ K. We note that these sub-band gaps are -- in addition to being in good quantitative agreement with experimental measurements \cite{Wagener_JAP_2014} -- close to the optimum values associated with an IBSC implemented using a GaAs host matrix. Specifically, for GaAs (which has $E_{g} = 1.42$ eV at $T = 300$ K) the optimum sub-band gaps $E_{\scalebox{0.6}{\textrm{L}}} = 0.45$ eV and $E_{\scalebox{0.6}{\textrm{H}}} = 0.97$ eV correspond to a detailed balance efficiency limit of $\approx 58$\% under concentrated illumination, close to the overall efficiency limit for an ideal IBSC. However, for a given host matrix band gap $E_{g}$ the detailed balance efficiency is a sensitive function of the sub-band gaps $E_{\scalebox{0.6}{\textrm{L}}}$ and $E_{\scalebox{0.6}{\textrm{H}}}$, reducing rapidly as the sub-band gaps are detuned from their optimum values. \cite{Wang_IETO_2014} It is therefore desirable to engineer the band structure such that the sub-band gaps are as close as possible to their optimum energies. The calculated value of $E_{\scalebox{0.6}{\textrm{L}}}$ ($E_{\scalebox{0.6}{\textrm{H}}}$) for the exemplar GaSb/GaAs QR considered above is only $\approx 50$ meV higher (lower) than the optimum value, suggesting that minor changes in morphology will be sufficient to realise optimum sub-band gaps.

In order to identify optimised QR morphologies we therefore proceed by repeating our analysis as a function of (i) QR height $h$, and (ii) QR Sb composition $x$, for QRs having the same inner and outer radii $a_{1} = 5$ nm and $a_{2} = 11.5$ nm considered above. We note that we have chosen to vary the QR height rather than in-plane dimensions since (i) real GaAs$_{1-x}$Sb$_{x}$/GaAs QRs typically have low aspect ratios $\frac{ h }{ 2a_{2} } \approx 0.1$, (ii) the confinement energies in such low aspect ratio structures, and hence the sub-band gaps, are dominated by confinement along the [001] direction (cf.~Eq.~\eqref{eq:quantum_ring_ground_state_energy}), and (iii) characterisation of epitaxially grown GaAs$_{1-x}$Sb$_{x}$/GaAs QRs has revealed significantly larger \textit{relative} variations in $h$ than in $a_{1}$ or $a_{2}$. The results of these calculations are summarised in Figs.~\ref{fig:numerical}(e) and~\ref{fig:numerical}(f), which respectively show the sub-band gap energy $E_{\scalebox{0.7}{\textrm{L}}}$ calculated based on the analytical and numerical models. Results are shown for GaAs$_{1-x}$Sb$_{x}$/GaAs QRs having Sb compositions $x = 100$, 90, 80 and 70\% using red, green, blue and orange closed circles, respectively.

Note that since the analytical model assumes a confining potential of infinite depth, it is not capable of directly predicting $E_{\scalebox{0.7}{\textrm{L}}}$ (which is obtained from the full numerical calculations as the difference $E_{h1} - E_{\scalebox{0.7}{\textrm{VB}}}$ in energy between the highest energy bound hole state and host matrix VB edge). In order to estimate $E_{\scalebox{0.7}{\textrm{L}}}$ using the analytical model we have therefore (i) used Eqs.~\eqref{eq:quantum_ring_transcendental} and~\eqref{eq:quantum_ring_ground_state_energy} to calculate the confinement energy $\Delta E_{h1}$ associated with the HH ground state in a QR of infinite potential depth, and (ii) used the (maximum) HH band offset $\Delta E_{\scalebox{0.7}{\textrm{HH}}}$ extracted from a full numerical calculation to compute $E_{\scalebox{0.7}{\textrm{L}}} = \Delta E_{\scalebox{0.7}{\textrm{HH}}} - \Delta E_{h1}$. Due to its assumption of an infinitely deep confining potential, the analytical approach naturally overestimates $\Delta E_{h1}$ -- in particular the contribution associated with confinement along [001] -- and hence underestimates $E_{\scalebox{0.7}{\textrm{L}}}$ for short QRs having $h \lesssim 3$ nm. Nonetheless, we observe that the analytical model accurately captures the key trends observed in the results of the full numerical calculations, and provides reasonably good quantitative agreement with the results of the numerical calculations for $h \gtrsim 3$ nm, across the full range of Sb compositions considered.

Examining Fig.~\ref{fig:numerical}(f), the results of our numerical calculations suggest that optimum IBSC sub-band gaps are obtained in short GaSb/GaAs QRs having $h \approx 2$ nm, which lies well within the range of heights observed in real structures. \cite{Khan_EMC_2016} By reducing the QR Sb composition $x$ at fixed $h$ to form alloyed GaAs$_{1-x}$Sb$_{x}$/GaAs QRs, we find that $E_{\scalebox{0.7}{\textrm{L}}}$ increases by $\approx 45$ -- 50 meV for each 10\% reduction in $x$. Therefore, in QRs having reduced Sb composition -- due, e.g., to interfacial Sb-As intermixing \cite{Timm_JVSTB_2008,Carrington_PB_2012} -- the QR height should be slightly increased in order to restore $E_{\scalebox{0.7}{\textrm{L}}}$ to its optimum energy. In practice, this constitutes reducing the confinement energy $\Delta E_{h1}$ so as to maintain fixed hole ionisation energy in response to the reduction in $\Delta E_{\scalebox{0.7}{\textrm{HH}}}$ associated with a reduction in $x$. For Sb compositions $x \gtrsim 75$\% our calculations indicate that an optimum sub-band gap $E_{\scalebox{0.7}{\textrm{L}}} = 0.45$ eV is maintained for an $\approx 1$ nm increase in $h$ in response to each 10\% reduction in $x$. For Sb compositions $x \lesssim 75$\% our calculations indicate that QR heights $h \gtrsim 5$ nm are required to obtain optimised electronic properties. Such heights are outside the range typically obtained via epitaxial growth, suggesting that only epitaxial growth of GaAs$_{1-x}$Sb$_{x}$/GaAs QRs having $x \gtrsim 75$\% is likely to produce heterostructures possessing electronic properties which are well-suited to IBSC applications.

We note, however, that our analysis of the electronic properties here has focused solely on idealised QR structures, possessing exact cylindrical shape and uniform alloy composition throughout the GaAs$_{1-x}$Sb$_{x}$ region. The presence of Sb-As intermixing at the ring-barrier interface in real GaAs$_{1-x}$Sb$_{x}$/GaAs QRs \cite{Timm_JVSTB_2008,Carrington_PB_2012} results in a non-uniform alloy composition, which may modify the strain fields and confining potentials compared to those considered in our exploratory calculations here. Overall, we expect the resulting modifications of the electronic properties to be quantitative rather than qualitative in nature compared to those described above for ideal QRs. Of more importance from the perspective of designing real IBSC devices is to build upon our initial analysis here by undertaking theoretical investigations of the optical properties of, and radiative and non-radiative losses in, GaAs$_{1-x}$Sb$_{x}$/GaAs QRs and vertical QR stacks.


\section{Conclusions}
\label{sec:conclusions}


In summary, we have presented a theoretical analysis of the electronic properties of type-II GaAs$_{1-x}$Sb$_{x}$/GaAs QRs based on a combined analytical and numerical approach, and identified optimised combinations of QR morphology and alloy composition for the realisation of hole-based IBSCs -- in which the IB is formed by the highest energy bound hole state in the QR -- offering the maximum theoretical efficiency available via inclusion of an IB in a GaAs matrix.


Analytically, we presented the solution of the time-independent Schr\"{o}dinger equation for a cylindrical QR of infinite potential depth and derived a transcendental equation which must be satisfied by bound QR eigenstates. The relationship to the solution of the well-known problem of the cylindrical QD was described, and it was demonstrated that (i) the QR eigenstates evolve smoothly from those of the QD, and (ii) the convergence properties of the QR ground state allow the confinement energy to be estimated straightforwardly, and to high accuracy, for realistic QRs having dimensions typical of those achieved via epitaxial growth. Our analytical analysis demonstrated that type-II GaAs$_{1-x}$Sb$_{x}$/GaAs QRs offer significant benefits from the perspective of band structure engineering for IBSC applications, allowing the IB energy in a hole-based IBSC to be tuned across a broad range via changes in QR morphology.

Numerically, we used multi-band \textbf{k}$\cdot$\textbf{p} calculations -- including full strain and piezoelectric effects -- to analyse the electronic properties of GaAs$_{1-x}$Sb$_{x}$/GaAs QRs, both as a function of QR dimensions and Sb composition. We further demonstrated that the nature of the carrier confinement in these heterostructures is ideally suited to IBSC applications. Strong hole localisation, with large ionisation energies in excess of 0.4 eV, can be expected to mitigate carrier leakage from the IB via thermionic emission. Additionally, the interplay between strain relaxation and type-II band alignment in these QRs was demonstrated to give rise to electron states which, in the plane of the QR, are strongly localised in the central barrier of the QR. The unusual nature of the carrier localisation in these heterostructures suggests the potential to engineer the trade-off between the electron-hole overlap (which mediates carrier generation via optical absorption) and the radiative lifetime for photo-generated electron-hole pairs (which mediates carrier loss via radiative recombination).

For inner and outer radii $a_{1} = 5$ nm and $a_{2} = 11.5$ nm, typical of epitaxially grown QRs, our calculations indicate that an optimum IB sub-band gap $E_{\scalebox{0.7}{\textrm{L}}} = E_{\scalebox{0.7}{\textrm{IB}}} - E_{\scalebox{0.7}{\textrm{VB}}}$ ($E_{\scalebox{0.7}{\textrm{H}}} = E_{\scalebox{0.7}{\textrm{CB}}} - E_{\scalebox{0.7}{\textrm{IB}}}$) of 0.45 eV (0.97 eV) can be obtained in GaSb/GaAs QRs having height $h \approx 2$ nm. For GaAs$_{1-x}$Sb$_{x}$/GaAs QRs having reduced Sb compositions $x$, our calculations indicate that decreases in $x$ can be compensated by slight increases in $h$ -- by $\approx 1$ nm for each 10\% reduction in $x$ -- in order to maintain optimum sub-band gap energies. QRs grown to these specifications have a detailed balance efficiency limit of $\approx 58$\% under concentrated illumination, close to the overall limit of 63.8\% for ideal IBSCs. Given the sensitivity of the theoretical IBSC efficiency to the IB energy, our analysis suggests that careful control of QR morphology provides a viable route to realising optimised heterostructures suitable for IBSC applications.

Our initial calculations here, however, were guided by detailed balance efficiency limits which include a number of assumptions -- e.g.~optimium absorption spectra, infinite carrier mobilities, absence of non-radiative carrier losses, etc.~-- which are not reflective of the conditions in real quantum-confined heterostructure. Further theoretical work is therefore required to quantify, and identify pathways towards simultaneously optimising the optical properties and mitigating losses (both radiative and non-radiative) in real GaAs$_{1-x}$Sb$_{x}$/GaAs QRs and vertical QR stacks, in order to identify rigorously optimised heterostructures for IBSC applications and to quantify photovoltaic efficiencies that can be realistically achieved using this novel platform.


Overall, our calculations highlight the suitability of type-II GaAs$_{1-x}$Sb$_{x}$/GaAs QRs for applications in hole-based IBSCs, and provide initial information regarding optimised combinations of QR alloy composition and morphology to guide the growth and fabrication of prototype QR-IBSC devices.


\section*{Acknowledgements}

This work was supported by the European Commission via the Marie Sk\l{}odowska-Curie Innovative Training Network PROMIS (project no.~641899), by the National University of Ireland (NUI; via the Post-Doctoral Fellowship in the Sciences, held by C.A.B.), and by Science Foundation Ireland (SFI; project no.~15/IA/3082). The authors thank Prof.~Anthony Krier and Dr.~Denise Montesdeoca (Lancaster University, U.K.) for useful discussions.






\end{document}